# Enhanced Coupling of Superconductivity and Evolution of Gap Structure in $CsV_3Sb_5$ through Ta Doping

Yiwen Li[1], Zhengyan Zhu[1], Qing Li[1], Yongze Ye[1], Zhiwei Wang[2,3,4*], Yugui Yao[2,3], and Hai-Hu Wen[1*]

[1]National Laboratory of Solid State Microstructures and Department of Physics, Collaborative Innovation Center of Advanced Microstructures, Nanjing University, Nanjing 210093, China

[2]Key Laboratory of Advanced Optoelectronic Quantum Architecture and Measurement, Ministry of Education, School of Physics, Beijing Institute of Technology, Beijing 100081, China

[3]Micronano Center, Beijing Key Lab of Nanophotonics and Ultrafine Optoelectronic Systems, Beijing Institute of Technology, Beijing 100081, China

[4]Material Science Center, Yangtze Delta Region Academy of Beijing Institute of Technology, Jiaxing, 314011, China

*Correspondence: zhiweiwang@bit.edu.cn; hhwen@nju.edu.cn

**Abstract**

In this study, we present a detailed investigation of kagome superconductors $CsV_3Sb_5$ single crystal and its Ta-doped variant, $Cs(V_{0.86}Ta_{0.14})_3Sb_5$, through specific heat measurements. Our results show a clear suppression of the charge density wave (CDW) and notable increase in the superconducting transition temperature ($T_c$) from 2.8 K to 4.6 K upon Ta doping. The electronic specific heat of the pristine $CsV_3Sb_5$ sample can be fitted with a model comprising an *s*-wave gap and a highly anisotropic extended *s*-wave gap, where the ratio $2\Delta/k_BT_c$ is smaller than the weak coupling limit of 3.5. For the doped sample $Cs(V_{0.86}Ta_{0.14})_3Sb_5$, it exhibits two isotropic *s*-wave gaps, yielding the larger gap of $2\Delta/k_BT_c = 5.04$, which indicates a significant enhancement in superconducting coupling. This evolution is attributed to the increased density of states (DOS) near the Fermi level released through the suppression of the CDW gap. Our results demonstrate enhanced superconducting coupling and variation of gap structure in $CsV_3Sb_5$ due to Ta doping.

# 1. Introduction

The newly discovered materials $A$V$_3$Sb$_5$ ($A$ = K, Rb, Cs) [1] have attracted significant attention due to their unique crystal structure. It consists of layers of V-Sb sheets, with adjacent layers separated by $A$-ion layers. In the V-Sb sheets, the V atoms form a kagome lattice. The kagome lattice is constructed from corner-sharing triangles, and this special geometry gives this system distinctive electronic structures. According to the angle-resolved photoemission spectroscopy (ARPES) experiments [2-11] and density functional theory (DFT) [2,12-14] calculations, $A$V$_3$Sb$_5$ exhibits a multiband structure with several energy bands crossing the Fermi level. Among them, V-3$d$ and Sb-5$p$ orbital electrons primarily contribute to the electronic states near the Fermi surface. There is an electron pocket near the Γ point in the Brillouin zone, which originates from the $p_z$ orbitals of Sb atoms. Near the M point, there are two van Hove singularities (VHSs) composed of multiple V-$d$ orbitals, including $d_{xy}$, $d_{x2-y2}$, and $d_{z2}$. Additionally, a Dirac cone near the K point arises from the $d_{xz}$ and $d_{yz}$ orbitals of V atoms. Such electronic states give rise to many intriguing phenomena for this system, such as CDW [15-18], anomalous Hall effect (AHE) [19-21], superconductivity [2,4,22,23], nontrivial band topology [4,24], charge-4e and charge-6e flux quantization [25,26], and residual Fermi arcs [27]. Among them, one of the most intriguing properties is the competition between CDW and superconductivity, and exploring their connections and interactions is crucial for unraveling the underlying physical mechanisms.

Doping is an important method for modulating the CDW and superconducting states. It can introduce positive or negative pressures that change the lattice constant and thus change the band structure, or it can introduce charge carriers or holes that affect the position of the Fermi surface. For example, Ti or Sn doping in CsV$_3$Sb$_5$ introduces hole-type carriers [28-30], leading to a double-dome superconductivity and progressive suppression of CDW. Cr or Mo doping introduces electron-type carriers [31,32], and Nb and Ta are iso-valent dopants [33,34]. They all exhibit different doping effects. By comparing the superconducting gap structure and coupling strength before and after doping, as well as the corresponding changes in the CDW state, and analyzing the differences in the Fermi surface and band structure, we can gain insights into the mechanisms and origins of these states.

In CsV$_3$Sb$_5$, the pairing symmetry and gap structure remain debated despite extensive research. Some evidence supports isotropic $s$-wave gaps. For instance, self-field critical current ($I_{c,sf}$) measurements are consistent with a model of a single conventional $s$-wave gap [35]. Additionally, techniques such as muon-spin relaxation/rotation ($\mu$SR) measurements [36,37], soft point-contact spectroscopy (SPCS) [38], specific heat [39], and tunnel diode oscillator (TDO) [39] also suggest the presence of two-gap $s$-wave superconductivity. On the contrary, other experiments have detected gap anisotropy. STM measurements [40-43] reveal V-shaped superconducting gaps, indicating significant anisotropy, while ARPES measurements [44] show large anisotropy (over 80%) in the superconducting gap on one of the Fermi pockets. Despite this discrepancy, there is a consensus that the multi-gap scenario is supported by most results in this multiband system. And currently, no direct evidence indicates the presence of nodes in the superconducting gap.

Specific heat measurement is a useful tool for probing low-energy quasiparticle excitations and providing bulk evidence of the order parameter and electronic states near the Fermi energy. In this study, we perform specific heat measurements on $CsV_3Sb_5$ and $Cs(V_{0.86}Ta_{0.14})_3Sb_5$ single crystals. The specific heat data in the two samples indicate that doping increases $T_c$ from 2.8 K to 4.6 K, and also reveal the suppression of the CDW order in $Cs(V_{0.86}Ta_{0.14})_3Sb_5$, confirming the competition between the superconducting state and the CDW order. For $Cs(V_{0.86}Ta_{0.14})_3Sb_5$, the low-temperature specific heat shows an upturn below 1 K, which is due to a Schottky anomaly caused by the Zeeman splitting of Ta ions. Analysis of the electronic specific heat for both samples reveals that the gap structure of the pristine sample can be fitted with an isotropic *s*-wave gap and a highly anisotropic extended *s*-wave gap (anisotropy parameter $\alpha = 0.8$), with $2\Delta/k_BT_c$ values of 2.98 and 2.49, consistent with the weak-coupling BCS theory. In contrast, for $Cs(V_{0.86}Ta_{0.14})_3Sb_5$, the gap structure turns into two isotropic *s*-wave gaps with $2\Delta/k_BT_c = 3.08$ and 5.04, where the larger gap corresponds to strong-coupling superconducting pairing. We attribute this enhancement to the doping-induced shift of the VHS away from the Fermi surface, which disrupts the Fermi surface nesting condition, suppresses the CDW gap, and thus increases the DOS at the Fermi level. This increased DOS significantly enhances superconducting pairing.

## 2. Results

Figures 1(a) and 1(b) display the magnetization data of $CsV_3Sb_5$ and $Cs(V_{0.86}Ta_{0.14})_3Sb_5$ measured in zero-field-cooling (ZFC) and field-cooling (FC) modes at magnetic fields of 3 Oe and 5 Oe, respectively. Both samples exhibit sharp superconducting transitions, indicative of their high quality. The diamagnetic transition occurs at 3.2 K for the pristine $CsV_3Sb_5$ and rises to 5.0 K after 14% Ta doping. The relationship between $T_c$ and Ta doping level in the series $Cs(V_{1-x}Ta_x)_3Sb_5$ is detailed in Ref. [45], showing that $T_c$ increases with the doping level until reaching the doping limit of 16% Ta.

Ta doping also affects the CDW behavior in $CsV_3Sb_5$. In pristine $CsV_3Sb_5$, an obvious CDW transition is observed at 94 K, as demonstrated by resistivity, magnetization, and specific heat measurements [2]. With Ta doping, resistivity measurements show that CDW gradually weakens and is eventually suppressed beyond 10% Ta doping [45]. In our study, we focus on $Cs(V_{0.86}Ta_{0.14})_3Sb_5$. Figure 1(c) presents specific heat data of $Cs(V_{0.86}Ta_{0.14})_3Sb_5$ from 2 K to 175 K. It is found that no detectable phase transition is observed in this range, which further confirms the suppression of the CDW order in terms of specific heat.

Then we measure the specific heat of the pristine $CsV_3Sb_5$ and doped sample $Cs(V_{0.86}Ta_{0.14})_3Sb_5$ down to 0.4 K. The specific heat of a material consists of phonon, electronic, and possibly magnetic contributions. The most valuable information, such as gap structure and pairing symmetry can be inferred from the electronic specific heat. By comparing the gap structures between undoped and doped samples and combining them with the Fermi surfaces and energy band structures observed by ARPES, we will see how Ta doping affects the gap structure, and how the energy bands near the Fermi surface influence the contribution to the superconducting energy gaps.

The specific heat data for $CsV_3Sb_5$ and $Cs(V_{0.86}Ta_{0.14})_3Sb_5$, plotted as $C/T$ versus $T$, are shown in Figs. 2(a) and 2(b). The square symbols represent the data at 0 T, which exhibit evident specific heat jumps at 2.8 K and 4.6 K, respectively. To suppress superconductivity, magnetic fields of 2 T and 4 T are applied to the two samples, as shown by the circle symbols. No superconducting signals are observed above 0.4 K for either sample under these fields.

To extract the electronic specific heat, we use the Debye model to approximate phonon contribution and subtract it from the total specific heat. In this analysis, the normal state specific heat is given by

$$C = \gamma_n T + \beta T^3 + \delta T^5$$

where the first term represents the electronic contributions and the latter two terms account for the phonon part. Here, $\gamma_n$, $\beta$, and $\delta$ are the temperature-independent fitting parameters. Sommerfeld constant $\gamma_n$ describes the DOS near the Fermi surface for the normal state, and the corresponding values are shown in Table I. The solid lines in Figs. 2(a) and 2(b) show the fitted results of the normal state specific heat, which closely match the experimental normal state data under magnetic fields. The inset of Fig. 2(b) shows the specific heat of $Cs(V_{0.86}Ta_{0.14})_3Sb_5$ under various magnetic fields. After subtracting the phonon component from the total specific heat, for $CsV_3Sb_5$, we obtain the electronic specific heat curves at 0 T in Fig. 2(c), while for $Cs(V_{0.86}Ta_{0.14})_3Sb_5$, the specific heat curves exhibit low-temperature upturns, as presented in Fig. 2(d). This upturn indicates that, in addition to the expected electronic contributions, there are significant magnetic contributions affecting the specific heat. The observed anomaly at low temperatures is attributed to the Schottky anomaly, which will be further discussed in the following part.

For clarification, the open symbols in Figs. 2(b) and 2(d) clearly deviating the trend just below the specific heat anomaly may be induced the poor temperature controlling since the temperature is close to the boiling point of liquid helium (LHe).

Since we have obtained the electronic specific heat of both samples, we can use the BCS model to describe it with the following expression:

$$\gamma_e = \frac{4N(E_F)}{k_B T^3} \int_0^{+\infty} \int_0^{2\pi} \frac{e^{\xi/k_B T}}{(1+e^{\xi/k_B T})^2} \left(\epsilon^2 + \Delta^2(\theta,T) - \frac{T}{2}\frac{d\Delta^2(T,\theta)}{dT}\right) d\theta d\epsilon$$

where $\gamma_e$ is the electronic specific heat coefficient, $\xi = \sqrt{\epsilon^2 + \Delta^2(T,\theta)}$, and $\epsilon = \frac{\hbar^2 k^2}{2m} - E_F$. The gap function $\Delta(T,\theta)$ is the gap function varies with different gap structures. For instance, a single $s$-wave gap is represented as $\Delta(T,\theta) = \Delta_0(T)$, a single $d$-wave gap as $\Delta(T,\theta) = \Delta_0(T)\cos 2\theta$, and a single extended $s$-wave as $\Delta(T,\theta) = \Delta_0(T)(1+\alpha\cos 2\theta)$, where $\alpha$ indicates anisotropy. If $\alpha = 1$, the gap function has accidental nodes, while $\alpha \approx 0$ suggests minimal anisotropy, leading to an isotropic $s$-wave gap.

In Fig. 3, we apply the BCS formula to the pristine CsV$_3$Sb$_5$. We first try to fit the specific heat with a one-gap model, as depicted in Fig. 3(a) and 3(b). it is found that neither a single isotropic $s$-wave gap nor a single extended $s$-wave gap fits the data, while a single $d$-wave gap appears to fit the data nicely. Despite this, the single $d$-wave model still has several issues. First, the sample does not exhibit any residual zero-temperature specific heat coefficient $\gamma_0$. Usually, for a $d$-wave superconductor, the existence of gap nodes allows to easily excite quasiparticles at low temperatures, which will result in a residual $\gamma_0$. Second, a recent ARPES experiment [44], which can directly measure the superconducting gap in the momentum space, does not provide evidence for a $d$-wave gap. Besides, according to the previous ARPES and DFT studies [2-14], the Fermi surface of CsV$_3$Sb$_5$ includes one circular sheet formed by Sb $5p$ electrons, and one triangle and one hexagonal sheet formed by V $3d$ electrons. This multiband structure will probably result in the multigap feature. In agreement with this, experiments such as specific heat [39], TDO [39], $\mu$SR [36,37], and ARPES [44] prove the multi-gap feature in this sample. Therefore, a single $d$-wave gap model may not be able to explain the specific heat with complicated band structure together with multiple components. As a result, we need to consider a two-gap model. Figure 3(c) presents a model with two isotropic $s$-wave gaps, which roughly fits the data with a larger gap $\Delta_1 = 0.425$ meV and a smaller gap $\Delta_2 = 0.145$ meV. The smaller gap is necessary for the fit at very low temperatures. However, above 2 K, there is a slight discrepancy between the fitted curve and the experimental data, with the fitted curve being slightly higher than the experimental data. To achieve a perfect match in this region, according to the principle of entropy conservation, the fitted curve below 0.4 K needs to decrease more slowly. This implies that a smaller gap minimum is needed. Thus, a fitting model with an isotropic $s$-wave gap and an extended $s$-wave gap emerges, as shown in Fig. 3(d). This model achieves a perfect fit, with an isotropic $s$-wave gap of 0.36 meV and an extended $s$-wave gap of 0.30 meV with an anisotropy parameter $\alpha = 0.8$.

The overall analysis of the three fitting methods suggests that the gap structure of CsV$_3$Sb$_5$ must include either a very small energy gap (~0.145 meV) or the presence of gap minima. Ultimately, since the two $s$-wave gap model is not as perfect as the other two models and the $d$-wave fit has several issues described above, we conclude that the gap structure in CsV$_3$Sb$_5$ includes an $s$-wave gap and a highly anisotropic extended $s$-wave gap. To conclusively determine the gap structure, the specific heat data at the lowest temperatures are crucial. Fortunately, a recent ARPES experiment [44], which detects a superconducting gap with an anisotropy over 80%, agrees with our results.

For the doped sample Cs(V$_{0.86}$Ta$_{0.14}$)$_3$Sb$_5$, after subtracting the phonon contribution from the total specific heat, the data reveals an upturn below 1 K, as shown in Fig. 2(d). This upturn is likely due to a Schottky anomaly, which complicates the BCS fitting process, as this anomaly needs to be ruled out for fitting.

Observing the 0 T curve in Fig. 2(d), the value of the specific heat coefficient is almost zero at 1 K, and the upturn appears below this temperature. This indicates that the specific heat below 1 K is mainly from the Schottky anomaly, while the specific heat above 1 K is mainly from the electronic state. Therefore, we only consider the data above 1 K for BCS fitting, as shown in Fig. 4.

The BCS fitting results of Cs(V$_{0.86}$Ta$_{0.14}$)$_3$Sb$_5$ are very different from those of the pristine CsV$_3$Sb$_5$. For one-gap models, neither a single $s$-wave nor a $d$-wave gap can fit the data, while if introducing a slight anisotropy of $\alpha = 0.3$ to the $s$-wave gap, it can fit the curve well. Considering the multi-gap characteristics of the pristine sample, we also apply a two-gap model for the Ta-doped sample. As shown in Fig. 4(c), the model of an $s$-wave gap and a $d$-wave gap cannot fit the data. In contrast, Fig. 4(d) shows the fit of two isotropic $s$-wave gaps with gap values of 0.61 meV and 1.0 meV, which perfectly matches the experimental curve. This result also agrees with the ARPES data on the same sample, where they directly observe nearly isotropic superconducting gaps in the momentum space [33].

Overall, a single extended $s$-wave gap with small anisotropy or a model of two isotropic $s$-wave gaps can describe the electronic specific heat of the doped sample. Compared to the pristine sample, there is a significant reduction in gap anisotropy in the doped sample. Finally, considering the multiband feature of the system, also in agreement with the ARPES data [33], we argue that the model of two isotropic $s$-wave gaps is better suited for Cs(V$_{0.86}$Ta$_{0.14}$)$_3$Sb$_5$.

We now discuss the low-temperature upturns in the specific heat of Cs(V$_{0.86}$Ta$_{0.14}$)$_3$Sb$_5$. In Fig. 5(a), we present the specific heat data after subtracting the phonon contributions from 0 T to 1 T for Cs(V$_{0.86}$Ta$_{0.14}$)$_3$Sb$_5$. All the data exhibit upturns at low temperatures, which means that the curves contain both electronic specific heat and magnetic contributions. In the last part, we have applied a two-$s$-wave-gap model to the electronic specific heat coefficient at 0 T. Therefore, for data under different fields, we continue to use the same model for temperatures above 1 K. Based on entropy conservation, the best fits are shown by the solid lines in Fig. 5(a). The fitting results for the data under magnetic fields become less accurate, which is due to the low-temperature anomaly shifting to higher temperatures with increasing field, introducing additional contributions to the electronic specific heat.

After subtracting the electronic contributions obtained from the BCS fit, the magnetic specific heat coefficients are shown in Fig. 5(b), which we believe is the two-level Schottky anomaly. Two-level Schottky anomaly is caused by energy level splitting. In a crystal field or magnetic field, particles undergo energy level splitting, producing ground state and excited states. At very low temperatures, transitions between these levels are minimal. As the temperature increases, particles in the ground state have a higher probability of being excited to higher energy states, thus increasing the specific heat. At even higher temperatures, the probabilities of occupying each level become equal, and no further transitions occur, resulting in a peak or bump in the Schottky specific heat. As the magnetic field increases, the Schottky peak should be suppressed and the weight is pushed to higher temperatures. The data in Fig. 5(b) clearly behave in that way.

For the data at 0 T or at small magnetic fields, we do not observe the peak due to the Schottky anomaly because the peak should appear at a very low temperature. In present case, it may appear below 0.4 K. However, for the data at 0.5 T, 1 T, and 2 T, we observe the phenomenon that the peak moves to higher temperatures and its height decreases as the magnetic field increases, which is consistent with the Schottky anomaly. We fit the data at 0.5 T, 1 T, and 2 T using the two-level Schottky specific heat formula [46]:

$$C_{Sch}(T,H) = n\left(\frac{g\mu_B H}{k_B T}\right)^2 \frac{e^{\frac{g\mu_B H}{k_B T}}}{\left(1 + e^{\frac{g\mu_B H}{k_B T}}\right)^2}$$

Here, $n$ is the density of magnetic particles, $g$ is the Landé factor, $H$ is the effective magnetic field, and $\mu_B$ is the Bohr magneton. As shown by the solid lines in Fig. 5(b), this formula fits the data well. Since the Schottky anomaly appears only in the doped sample, we attribute it to the Zeeman splitting of Ta ions.

## 3. Discussion

Based on our analysis of the specific heat for $CsV_3Sb_5$ and $Cs(V_{0.86}Ta_{0.14})_3Sb_5$, we summarize the obtained parameters in Table 1, and next, we will have a discussion.

In the pristine $CsV_3Sb_5$, superconductivity coexists with the CDW order. The $T_c$ in this material is about 2.8 K, while the CDW transition occurs at 94 K. After doping with 14% Ta, $T_c$ increases to 4.6 K, and the CDW order is suppressed. This observation proves the competition between superconductivity and CDW in this system.

Besides, doping changes the gap structure of this material. Our BCS fittings of the specific heat show that the gap structure in the pristine sample can be described with the model of an isotropic *s*-wave gap and an extended *s*-wave gap with an anisotropy of 0.8. After Ta doping, the gap structure evolves into two isotropic *s*-wave gaps. This conclusion is consistent with the STM [43] and ARPES [44,33] experiments. According to the STM experiments [43], they detect a V-shape superconducting gap in the pristine $CsV_3Sb_5$, indicating the existence of significant gap minima and quasiparticle excitations. Then with increasing Ta doping, this V-shape gap turns into a U-shaped gap, which indicates that the gap is almost isotropic. Similarly, in the ARPES experiment [44,33], it is found that for $CsV_3Sb_5$, a gap with over 80% anisotropy exists on the hexagonal pocket ($\beta$ pocket) at the Γ point. Also, there are isotropic gaps on both the circular Fermi pocket ($\alpha$ pocket) at the Γ point and the triangular pocket ($\delta$ pocket) at the K point. For $Cs(V_{0.86}Ta_{0.14})_3Sb_5$, the anisotropic gap on the $\beta$ pocket evolves to become isotropic. Combing our BCS fit results with the ARPES analysis in $CsV_3Sb_5$, we propose that the extended *s*-wave gap with anisotropy of 0.8 corresponds to the gap on the $\beta$ pocket, which is contributed from V-3*d* orbital, while the isotropic *s*-wave gap might come from $\alpha$ or $\delta$ pocket.

Then we calculate the parameter of $2\Delta/k_B T_c$, which measures the pairing strength.

For CsV$_3$Sb$_5$, the two gaps yield $2\Delta/k_BT_c$ values of 2.98 and 2.49, which are smaller than the weak coupling limit of 3.5. For Cs(V$_{0.86}$Ta$_{0.14}$)$_3$Sb$_5$, the values of $2\Delta/k_BT_c$ turn to 5.04 and 3.08, corresponding to the gaps of 1.0 meV and 0.61 meV. Obviously, the large gap of 1.0 meV in the doped sample belongs to the category of strong coupling. This means that doping enhances the superconducting pairing. Meanwhile, we also find a significant increase in the Sommerfeld constant $\gamma_n$ from 22.2 mJ mol$^{-1}$ K$^{-2}$ to 34.1 mJ mol$^{-1}$ K$^{-2}$ with doping. These effects are related to the interplay between CDW and superconductivity. In the pristine CsV$_3$Sb$_5$, when the band filling approaches the VHS, the hexagonal Fermi surface of the kagome lattice is more likely to Fermi surface nesting. This nesting can lead to Fermi surface instability, opening a gap and entering a CDW state. The gap opening at CDW state will reduce the DOS near the Fermi surface dramatically. While doping changes this condition. For example, the ARPES experiment on Ti-doped sample Cs(V$_{1-x}$Ti$_x$)$_3$Sb$_5$ [47], shows that Ti dopant will push the VHS above the Fermi level, which destroys the nesting condition and suppresses the CDW state. Therefore, the DOS near the Fermi surface which was gapped by the CDW will be released and lead to an increased $\gamma_n$. These DOS near the Fermi surface will participate in the superconducting pairing, so that we find the enhanced coupling of superconductivity for the doped sample. Additionally, we calculate the parameter of $\Delta C/\gamma_nT_c$, which can also measure the pairing strength. The pristine CsV$_3$Sb$_5$ has a value of 0.96, consistent with weak coupling theory (smaller than 1.43), while Cs(V$_{0.86}$Ta$_{0.14}$)$_3$Sb$_5$ shows a value of 2.26, indicating strong pairing.

However, there is another point of view. An ARPES experiment on Ta doped sample CsV$_{3-x}$Ta$_x$Sb$_5$ (x about 0.4) observes that the VHS does not move above the Fermi surface but perfectly aligns with the Fermi level, creating a flat dispersion at the Fermi surface, which can greatly increase the DOS participating in superconducting pairing [11]. This could also be a reason for the enhanced superconducting pairing. Interestingly, if the VHS were precisely on the Fermi surface, the Fermi surface nesting condition would still stand, but CDW has indeed been suppressed, which contradicts the traditional view that CDW is induced by the nesting of the momentum zones of VHS.

## 4. Conclusion

In summary, we have carried out specific heat measurements on CsV$_3$Sb$_5$ single crystals and its Ta doping sample Cs(V$_{0.86}$Ta$_{0.14}$)$_3$Sb$_5$. We find an increase in $T_c$ and a suppression in CDW through Ta doping. Besides, for Cs(V$_{0.86}$Ta$_{0.14}$)$_3$Sb$_5$, low-temperature specific heat displays a Schottky anomaly which may be caused by the Zeeman splitting of Ta ions. BCS fittings of the electronic specific heat for both samples reveal a change in the gap structure: from one isotropic *s*-wave gap plus one anisotropic *s*-wave gap (with anisotropy of 0.8) to two isotropic *s*-wave gaps. Furthermore, calculating $2\Delta/k_BT_c$, shows an enhanced superconducting coupling with doping. Based on the band structure analysis, this enhanced superconductivity is attributed to the increase of DOS near the Fermi surface which is influenced by the VHS.

**Materials and methods**

The single crystals $CsV_3Sb_5$ and $Cs(V_{0.86}Ta_{0.14})_3Sb_5$ single were grown by the self-flux method [1,2,33]. The magnetization measurements were carried out with a superconducting quantum interference device (SQUID-VSM, Quantum Design). The specific heat measurements were carried out with a physical property measurement system (PPMS, Quantum Design) by thermal-relaxation method.

**Data availability**

The data that support the findings of this study are available in the paper. Additional data are available from the corresponding authors upon reasonable request.

**Acknowledgments**

This work was supported by the National Key R&D Program of China (No. 2022YFA1403201, No. 2022YFA1403400, and No. 2020YFA0308800), the National Natural Science Foundation of China (No. 11927809, 12061131001, No. 11974171, No. 92065109, and No. 12204231), the Funda- mental Research Funds for the Central Universities (No. 020414380208). Z.W. thanks the Analysis & Testing Center at BIT for assistance in facility support.

**Declarations**

**Competing interests**

The authors declare no competing interests.

**Author contributions**

LY, ZZ, YY, and WHH measured and analyzed the data. LQ, WZ, and YY grew the samples. LY and WHH wrote the manuscript. All the authors read and approved the final manuscript.

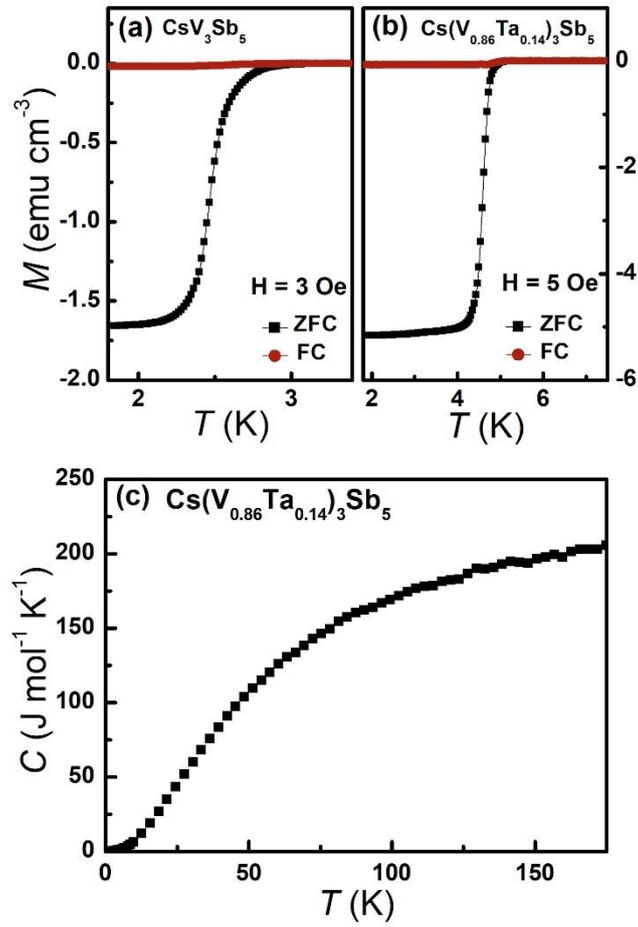

**Figure 1** Temperature dependence of magnetization for (**a**) CsV$_3$Sb$_5$ and (**b**) Cs(V$_{0.86}$Ta$_{0.14}$)$_3$Sb$_5$ in ZFC and FC modes. (**c**) Temperature dependence of specific heat for Cs(V$_{0.86}$Ta$_{0.14}$)$_3$Sb$_5$ from 2 K to 175 K

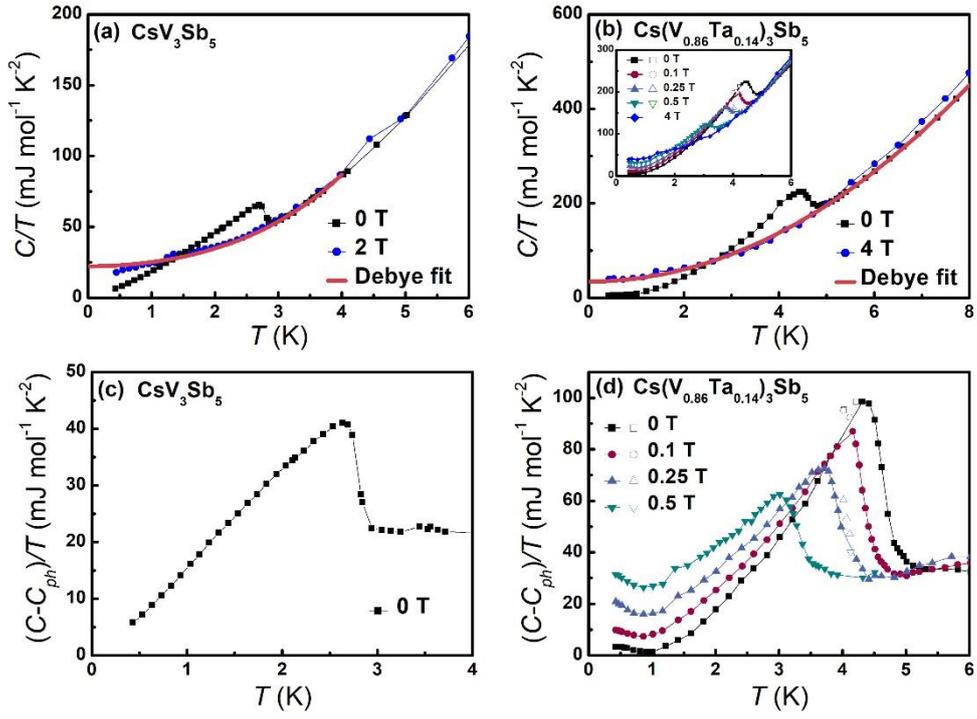

**Figure 2** Temperature dependence of specific heat, plotted as *C/T* versus *T*, for (**a**) CsV$_3$Sb$_5$ and (**b**) Cs(V$_{0.86}$Ta$_{0.14}$)$_3$Sb$_5$. The solid lines represent the fitted data according to the Debye model (see text). The inset of (**b**) shows the specific heat under various magnetic fields. Temperature dependence of specific heat after subtracting their phonon contributions, for (**c**) CsV$_3$Sb$_5$ and (**d**) Cs(V$_{0.86}$Ta$_{0.14}$)$_3$Sb$_5$. The open symbols in (**b**) and (**d**) are the error points around 4 K.

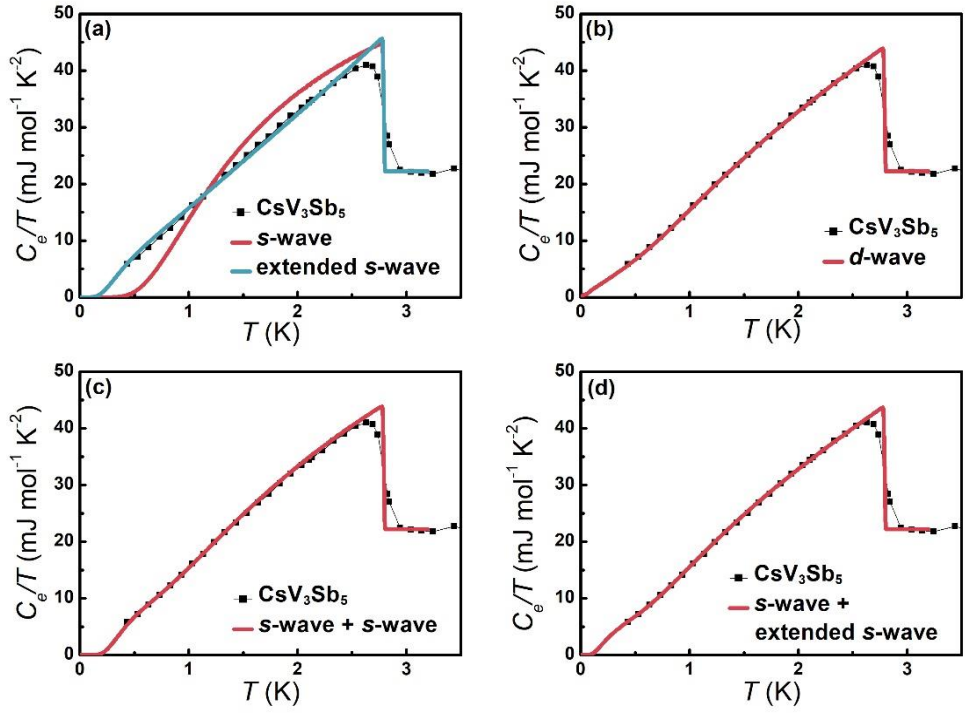

**Figure 3** Temperature dependence of electronic specific heat coefficient for $CsV_3Sb_5$ and the BCS fit results with (**a**), (**b**) one-gap model and (**c**), (**d**) two-gap model (see text)

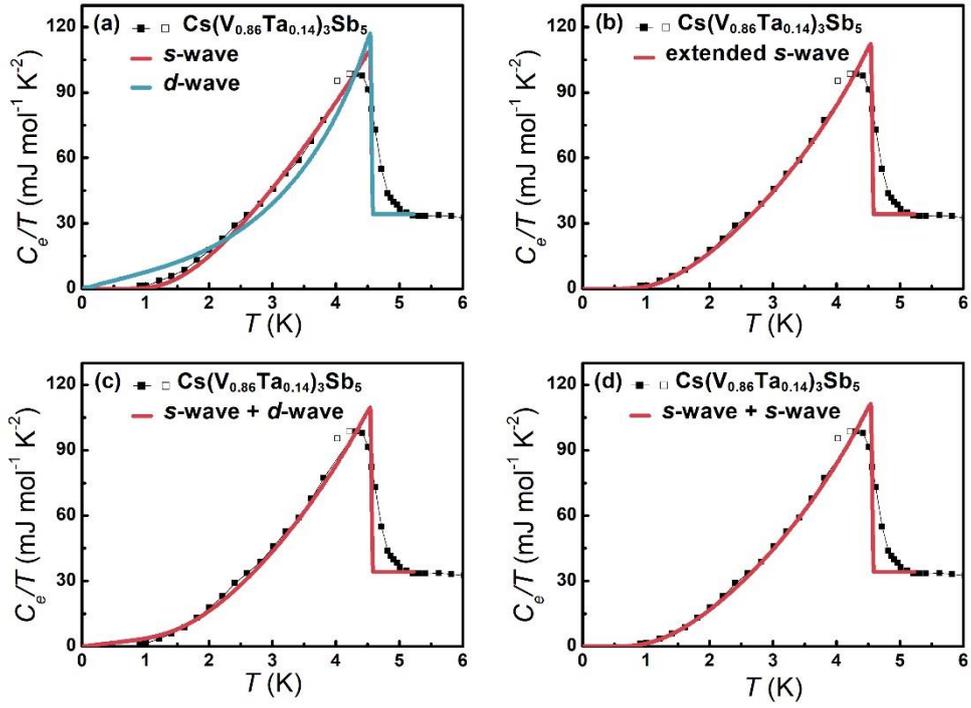

**Figure 4** Temperature dependence of electronic specific heat coefficient above 1 K for Cs(V$_{0.86}$Ta$_{0.14}$)$_3$Sb$_5$ and the BCS fit results with (**a**), (**b**) one-gap model and (**c**), (**d**) two-gap model (see text). The two open symbols are the error points.

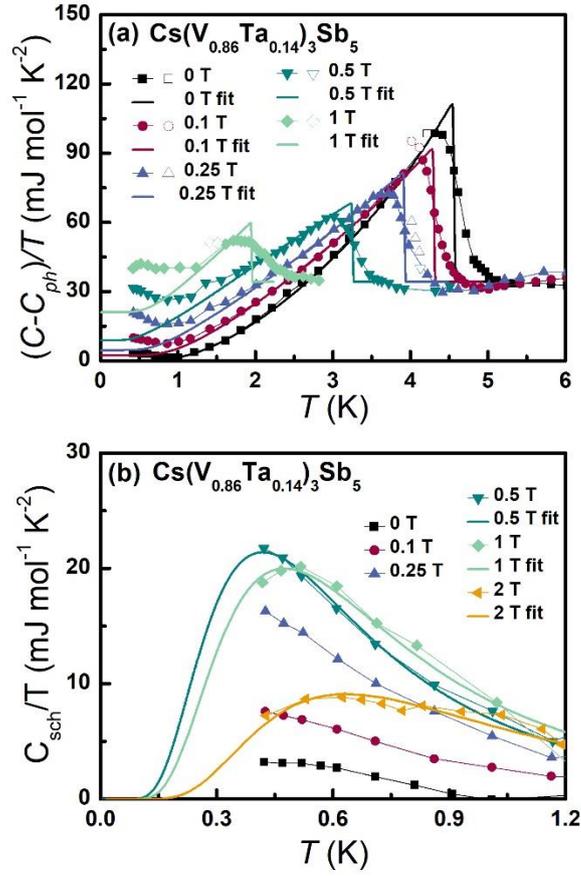

**Figure 5** (**a**) Temperature dependence of the specific heat after subtracting phonon contributions under different magnetic fields for Cs(V$_{0.86}$Ta$_{0.14}$)$_3$Sb$_5$. Solid lines are the BCS fit results. The open symbols are the error points. (**b**) The magnetic specific heat coefficient of Cs(V$_{0.86}$Ta$_{0.14}$)$_3$Sb$_5$ under different fields. Solid lines are their Schottky fits (see text).

|  | $CsV_3Sb_5$ | $Cs(V_{0.86}Ta_{0.14})_3Sb_5$ |
|---|---|---|
| $T_c$ | 2.8 K | 4.6 K |
| CDW | √ | × |
| Gap structure | $s$-wave ($\Delta$ = 0.36 meV) + extended $s$-wave ($\Delta$ = 0.30 meV, $\alpha$ = 0.8) | $s$-wave ($\Delta$ = 1.0 meV) + $s$-wave ($\Delta$ = 0.61 meV) |
| $2\Delta/k_B T_c$ | 2.98/2.49 | 5.04/3.08 |
| $\gamma_n$ | 22.2 mJ mol$^{-1}$ K$^{-2}$ | 34.1 mJ mol$^{-1}$ K$^{-2}$ |
| $\Delta C/\gamma_n T_c$ | 0.96 | 2.26 |

**Table 1** Several parameters of $CsV_3Sb_5$ and $Cs(V_{0.86}Ta_{0.14})_3Sb_5$ for comparison